\documentclass[amsmath,11pt]{article}

\usepackage{amsfonts,amsmath,amssymb,graphics,epsfig}

\date{\empty}

\begin{document}

\author{Klaountia Pasmatsiou${}^{1,2}$, Christos G. Tsagas${}^{2}$ and John D. Barrow$^{3}$\\ {\small ${}^1$CERCA, Department of Physics, Case Western Reserve University}\\ {\small 10900 Euclid Avenue, Cleveland, OH 44106, USA}\\ {\small ${}^2$Section of Astrophysics, Astronomy and Mechanics, Department of Physics}\\ {\small Aristotle University of Thessaloniki, Thessaloniki 54124, Greece}\\ {\small ${}^3$DAMTP, Centre for Mathematical Sciences, University of Cambridge}\\ {\small Wilberforce Road, Cambridge~CB3~0WA, UK}}

\title{\bf Kinematics of Einstein-Cartan universes}

\maketitle

\begin{abstract}
We analyse the kinematics of cosmological spacetimes with nonzero torsion, in the framework of the classical Einstein-Cartan gravity. After a brief introduction to the basic features of spaces with non-vanishing torsion, we consider a family of observers moving along timelike worldlines and focus on their kinematic behaviour. In so doing, we isolate the irreducible variables monitoring the observers' motion and derive their evolution formulae and associated constraint equations. Our aim is to identify the effects of spacetime torsion, and the changes they introduce into the kinematics of the standard, torsion-free, cosmological models. We employ a fully geometrical approach, imposing no restrictions on the material content, or any a priori couplings between torsion and spin. Also, we do not apply the familiar splitting of the equations, into a purely Riemannian component plus a torsion/spin part, at the start of our study, but only introduce it at the very end. With the general formulae at hand, we use the Einstein-Cartan field equations to incorporate explicitly the spin of the matter. The resulting formulae fully describe the kinematics of dynamical spacetimes within the framework of the Einstein-Cartan gravity, while in the special case of the so-called Weyssenhoff fluid, they recover results previously reported in the literature.
\end{abstract}

\section{Introduction}\label{sI}
General relativity accounts for the macroscopic distribution of matter. It is therefore reasonable to view Einstein's theory as the macroscopic limit of a, still illusive, microphysical theory of gravity. The first steps towards such a theory were probably taken by \'Elie Cartan, who suggested that spacetime torsion could be used as the macroscopic manifestation of the intrinsic angular momentum of the matter~\cite{C}. Cartan's theory, however, was proposed before the discovery of the electron spin and this was perhaps one of the reasons his ideas went essentially unnoticed for some decades. It was probably not until the work of Kibble and Sciama, who laid down the foundations of $U_4$ theory, that the role of spacetime torsion in modern physics was appreciated~\cite{Ki}. Soon after that, a geometrical approach to the new theory was introduced as well~\cite{HK}. For a recently published collection of classic papers on the subject, with corresponding commentaries, we refer the reader to~\cite{BH}.

The Einstein-Cartan gravity, or the Einstein-Cartan-Kibble-Sciama (ECKS) theory as it is sometimes also referred to, is a viable description of the gravitational field that introduces an additional (rotational) degree of freedom to the spacetime fabric. The latter is carried by the non-Riemannian (the torsional) component of the affine connection and it is macroscopically related to the intrinsic angular momentum (spin) of the matter. By coupling the energy density and the spin of the matter to the metric and the torsion tensors respectively and by treating them as independent variables, the Einstein-Cartan gravity provides the simplest classical extension of general relativity. The predictions of the theory are essentially indistinguishable from those of general relativity even at nuclear densities, with departures appearing only at extremely high densities, like those anticipated in black-hole interiors and the very early universe. In these environments, the coupling between spin and torsion leads to a repulsive gravitational ``force'', which could (in principle at least) prevent the formation of singularities (e.g.~in Friedmann-Robertson-Walker (FRW) cosmologies, or in Bianchi-type models~\cite{K1}-\cite{SH}).

Since its reemergence in the late 1950s, the Einstein-Cartan gravity has gone though several phases of renewed interest, motivated by the ongoing effort to extend, compare and possibly link general relativity to the theories of the microphysical interactions. The kinematics of the theory have been investigated by several authors in an attempt to establish the effects of torsion and spin, primarily (though not exclusively) on the mean expansion of the Einstein-Cartan universes~\cite{T}-\cite{BHL}. Almost all of the approaches start by splitting their equations into a purely Riemannian (general relativistic) part plus a component conveying the effects torsion and spin. Also, matter is usually represented by the so-called Weyssenhoff fluid, namely an ideal medium with a specific ``equation of state'' for the spin density~\cite{WB}. Here, we do not apply such a decomposition until the very end of the study. Instead, our kinematic equations (both the propagation formulae and the constraints) are derived in successive steps. First by incorporating the effects of spacetime torsion and then those of the matter spin. Moreover, we do not impose any simplifying symmetries (e.g.~homogeneity or isotropy) and in so doing we provide a complete 1+3 covariant kinematic description of spacetimes with nonzero torsion and spin, along the lines of the classic (torsion-free) study of~\cite{E}.

We start by assuming Riemann-Cartan geometry and without making any a priori assumptions on the nature of the gravitational interaction, or on the relation between torsion and spin. These are specified in a subsequent step by means of the Einstein-Cartan and the Cartan field equations, though still without specifying the nature of the matter fields. All these mean that the resulting two sets of formulae (which are new -- to the best of our knowledge), apply to a general Riemann-Cartan spacetime and then to an Einstein-Cartan universe with arbitrary matter, respectively. The latter is specified at the very end of our study, where we also derive the Raychaudhuri equation of a Weyssenhoff fluid, thus allowing for a comparison with the existing literature. Our results confirm those of earlier studies, namely that the spin of the Weyssenhoff medium can inhibit (perhaps even reverse) its gravitational collapse, or assist its volume expansion. We also demonstrate that the aforementioned effects come into play through spin-induced changes in the rotational behaviour of the spacetime, rather than from the spin's contribution to the local gravitational field. Alternatively, that is for media with non-vanishing spin vector, the macroscopic effect of the particles' intrinsic angular momentum to the associated Raychaudhuri equation also depends on the ``tilt angle'' between the spin vector and the 4-velocity of the fluid.

After a brief introduction of the concept of torsion, we outline how the latter alters key features of Riemannian spaces, such as the operation of covariant differentiation and the interpretation of the geodesic lines. In the next two sections, we discuss the basic geometric properties of spaces with nonzero torsion, before proceeding to the so-called Riemann-Cartan spacetimes. Our starting point is the kinematics of timelike worldlines embedded in the aforementioned spacetimes. This takes place in section 5, where we also provide a direct comparison between the related Riemannian and Riemann-Cartan (irreducible) kinematics variables. These are defined by employing the so-called 1+3 covariant formalism, which facilitates a geometrical approach that combines mathematical compactness and clarity with physical transparency. Sections 6 and 7 derive the three evolution and the three constraint equations monitoring the kinematic behaviour of metric spaces with non-vanishing torsion. Our formulae are applied to matter fields with nonzero spin, by employing the Einstein-Cartan and the Cartan field equations, in section 8. There, we also consider a number of the special cases and among them that of the Weyssenhoff fluid and re-examine, following an alternative route, how its spin can affect the mean kinematics of the host spacetime.

\section{Spaces with torsion}\label{sST}
Riemannian geometry demands the symmetry of the affine connection, which means that space has zero torsion by default. Nevertheless, one could treat (classical) torsion as an independent variable/field, in addition to the metric, and thus ``replace'' the Riemannian spaces with their more general Riemann-Cartan counterparts.

\subsection{The contortion tensor}\label{ssCT}
Consider a general metric space with asymmetric affine connection $\Gamma^a{}_{bc}$. Demanding the invariance of the metric tensor under covariant differentiation, namely imposing the metricity condition $\nabla_cg_{ab}=0$, leads to the following expression for the connection
\begin{equation}
\Gamma^a{}_{bc}= \tilde{\Gamma}^a{}_{bc}+ K^a{}_{bc}\,.  \label{Gammas}
\end{equation}
Here, $\tilde{\Gamma}^a{}_{bc}$ are the Christoffel symbols of the associated Riemannian space and $K^a{}_{bc}$ the so-called contortion tensor.\footnote{Throughout this manuscript, tildas will indicate Riemannian variables related to the Christoffel symbols only. We also adopt a spacetime metric with signature ($-,+,+,+$) and set the speed of light equal to unity.} The latter is defined by
\begin{equation}
K^a{}_{bc}= S^a{}_{bc}+ S_{bc}{}^a+ S_{cb}{}^a= S^a{}_{bc}+ 2S_{(bc)}{}^a\,,  \label{cont}
\end{equation}
with $S^a{}_{bc}=\Gamma^a{}_{[bc]}$ representing Cartan's torsion tensor (determined by 24 independent components).\footnote{In the literature there are alternative definitions of the torsion and the contortion tensors. Here we follow those of~\cite{P1}, though there the metric signature is ($+,-,-,-$). Alternatively, one may define the torsion tensor as $S_{bc}{}^a=\Gamma^a{}_{[bc]}$ and the contortion tensor as $K_{ab}{}^c=S_{ab}{}^c-S_b{}^c{}_a+ S^c{}_{ab}= S_{ab}{}^c+2S^c{}_{(ab)}$~\cite{S}.} Geometrically speaking, the effect of space torsion is to prevent infinitesimal parallelograms from closing (e.g.~see~\cite{HvdHK}). Physically, torsion can provide a possible link between the spacetime geometry and the intrinsic angular momentum (i.e.~the spin) of the matter.

Staring from definition (\ref{cont}) and employing some straightforward algebra, one can show that the contortion tensor satisfies the symmetries
\begin{equation}
K_{abc}= K_{[ab]c}\,, \hspace{10mm} K_{a(bc)}= 2S_{(bc)a}\,, \hspace{10mm} K_{a[bc]}= S_{abc}  \label{contsym1}
\end{equation}
and
\begin{equation}
K_{(a|b|c)}= -2S_{(ac)b}\,, \hspace{10mm} K_{[a|b|c]}= -S_{bac}\,.  \label{contsym2}
\end{equation}
It follows that, in the special case of a fully antisymmetric torsion tensor (i.e.~when $S_{abc}=S_{[abc]}$), the contortion tensor reduces to $K_{abc}=S_{abc}$ and becomes totally skew as well (i.e.~$K_{abc}=K_{[abc]}$ -- see Eq.~(\ref{cont}) above). Definition (\ref{cont}), together with relation (\ref{contsym1}b), ensures that $\Gamma^a{}_{(bc)}=\tilde{\Gamma}^a{}_{bc}+ 2S_{(bc)}{}^a\neq \tilde{\Gamma}^a{}_{bc}$. In other words, the symmetric part of the general connection does not coincide with the Christoffel symbols of the corresponding (torsion-free) Riemannian space.

We finally note that expression (\ref{Gammas}) also guarantees the invariance of the metric tensor with respect to covariant differentiation in terms of the Levi-Civita connection (i.e.~the Christoffel symbols). In other words, in addition to $\nabla_cg_{ab}=0$, we have $\tilde{\nabla}_cg_{ab}=0$ as well.

\subsection{The torsion vector}\label{ssTV}
The antisymmetry of the torsion tensor translates into $S^a{}_{ab}= -S^a{}_{ba}$ and $S^a{}_b{}^b=0$. As a result, there is only one non-trivial contraction of $S^a{}_{bc}$, which defines the so-called torsion vector
\begin{equation}
S_a= S^b{}_{ab}= -S^b{}_{ba}\,.  \label{torsv}
\end{equation}
It follows that a totally antisymmetric torsion tensor is traceless with zero torsion vector by default. Given that the torsion tensor is trace-free  when the torsion vector vanishes and vice-versa, the ``modified'' torsion tensor
\begin{equation}
\mathfrak{S}^a{}_{bc}= S^a{}_{bc}+ {2\over3}\,\delta^a{}_{[b}S_{c]}\,,  \label{cS}
\end{equation}
is traceless by construction. The contractions of the contortion tensor follow directly from definitions (\ref{cont}) and (\ref{torsv}) and they are given by
\begin{equation}
K_{ab}{}^b= -2S_a\,, \hspace{10mm} K^b{}_{ab}= 2S_a \hspace{10mm} {\rm and} \hspace{10mm} K^b{}_{ba}= 0\,.  \label{Conttr}
\end{equation}
Clearly, a totally skew torsion tensor corresponds to a fully antisymmetric and traceless contortion tensor and vice versa.

\subsection{Autoparallel and geodesic curves}\label{ssAGCs}
In metric spaces with non-vanishing torsion, there are two types of preferred curves, namely the autoparallel and the geodesic curves. The former are the "straightest" lines and the latter are the lines of "extremum" (i.e.~minimum/maximum) length~\cite{HvdHK}. Both reduce to the familiar geodesic curves of the associated Riemannian space when the torsion is switched off.

Consider a curve with parametric equations $x^a=x^a(s)$, where $s$ is an affine parameter and $u^a={\rm d}x^a/{\rm d}s$ is the corresponding tangent vector. By definition the ``autoparallel'' equation is obtained after imposing the condition of parallel transport along the curve in question, namely by assuming that $u^b\nabla_bu^a= 0$. The latter immediately translates into the autoparallel equation
\begin{equation}
{{\rm d}^2x^a\over{\rm d}s^2}+ \Gamma^a{}_{bc}{{\rm d}x^b\over{\rm d}s}{{\rm d}x^c\over{\rm d}s}= 0\,.  \label{autop1}
\end{equation}
Note that only the symmetric part of the connection contributes to the right-hand side of the above, which is however torsion-dependent (see \S~\ref{ssCT} previously)

Geodesics are curves of extremal length. Since the distance (i.e.~the line element) between any two points depends only on the metric and not on the torsion, the geodesic equation reads
\begin{equation}
{{\rm d}^2x^a\over{\rm d}s^2}+ \tilde{\Gamma}^a{}_{bc}{{\rm d}x^b\over{\rm d}s}{{\rm d}x^c\over{\rm d}s}= 0\,.  \label{geod1}
\end{equation}
exactly as in the associated Riemannian space~\cite{HvdHK}.

Using definition (\ref{Gammas}), together with the symmetries of the contortion tensor (see Eq.~(\ref{contsym1}b) in \S~\ref{ssCT}), expression (\ref{autop1}) recasts into
\begin{equation}
{{\rm d}^2x^a\over{\rm d}s^2}+ \tilde{\Gamma}^a{}_{bc}{{\rm d}x^b\over{\rm d}s}{{\rm d}x^c\over{\rm d}s}+ 2S_{bc}{}^a{{\rm d}x^b\over{\rm d}s}{{\rm d}x^c\over{\rm d}s}= 0\,.  \label{autop2}
\end{equation}
In the absence of torsion, the above immediately reduces to Eq.~(\ref{geod1}). Moreover, in line with (\ref{autop2}), autoparallels and geodesics can coincide even for nonzero torsion, provided that $S_{(ab)}{}^c=0$. The latter ensures the total antisymmetry of the torsion tensor (i.e.~$S_{abc}=S_{[abc]}$), in which case the torsion vector vanishes identically (i.e.~$S_a=0$ -- see \S~\ref{ssTV} above).

\section{Curvature with torsion}\label{sCT}
Introducing an affine connection different from the Christoffel symbols, means that the geometry of the space is not entirely described by the metric. Instead, the Riemann-Cartan space has additional independent features that are encoded in the torsion/contortion tensor.

\subsection{The Riemann-Cartan tensor}\label{ssR-CT}
The curvature tensor of a general (not necessarily metric) space is obtained from the associated connection, in line with the familiar relation (e.g.~see~\cite{Pa})
\begin{equation}
R^a{}_{bcd}= \partial_c\Gamma^a{}_{bd}- \partial_d\Gamma^a{}_{bc}+ \Gamma^s{}_{bd}\Gamma^a{}_{sc}- \Gamma^s{}_{bc}\Gamma^a{}_{sd}\,.  \label{curvt1}
\end{equation}
In a metric space with non-vanishing torsion we have $\Gamma^a{}_{bc}=\tilde{\Gamma}^a{}_{bc}+K^a{}_{bc}$ (see Eq.~(\ref{Gammas}) earlier), which substituted into the right-hand side of the above provides the following expression for the Riemann-Cartan curvature tensor
\begin{equation}
R^a{}_{bcd}= \tilde{R}^a{}_{bcd}+ Q^a{}_{bcd}+ \tilde{\Gamma}^s{}_{bd}K^a{}_{sc}+ K^s{}_{bd}\tilde{\Gamma}^a{}_{sc}- \tilde{\Gamma}^s{}_{bc}K^a{}_{sd}- K^s{}_{bc}\tilde{\Gamma}^a{}_{sd}\,,  \label{curvt2}
\end{equation}
where
\begin{equation}
\tilde{R}^a{}_{bcd}= \partial_c\tilde{\Gamma}^a{}_{bd}- \partial_d\tilde{\Gamma}^a{}_{bc}+ \tilde{\Gamma}^s{}_{bd}\tilde{\Gamma}^a{}_{sc}- \tilde{\Gamma}^s{}_{bc}\tilde{\Gamma}^a{}_{sd}\,,  \label{tRt}
\end{equation}
is the associated (torsion-free) Riemann curvature tensor and
\begin{equation}
Q^a{}_{bcd}= \partial_cK^a{}_{bd}- \partial_dK^a{}_{bc}+ K^s{}_{bd}K^a{}_{sc}- K^s{}_{bc}K^a{}_{sd}\,.  \label{Qt}
\end{equation}
Given the close formalistic analogy between $\tilde{R}^a{}_{bcd}$ and $Q^a{}_{bcd}$, the latter may be seen as the purely-torsional counterpart of the Riemann curvature tensor. According to expressions (\ref{curvt2})-(\ref{Qt}), the curvature tensor of a general space with non-vanishing torsion decomposes into an exclusively Riemannian, a purely torsional and a mixed component.

Nonzero torsion means that the Riemann-Cartan curvature tensor no longer satisfies all the symmetries of its Riemannian counterpart. More specifically, definition (\ref{curvt2}) and the Ricci identities of a general space with torsion, ensure that $R_{abcd}= R_{[ab][dc]}$ (see footnote~4 in \S~\ref{sKE} below). In general, however, $R_{abcd}\neq R_{cdab}$ and $R^a{}_{[bcd]}\neq0$.

\subsection{The Ricci-Cartan tensor}\label{ssRicci-CT}
The symmetries of the Riemann-Cartan curvature tensor guarantee that the associated Ricci tensor ($R_{ab}=R^c{}_{acb}$) remains uniquely defined, despite the presence of torsion. On the other hand, we have $R_{abcd}\neq R_{cdab}$, which implies that the Ricci curvature tensor is not necessarily symmetric (i.e.~$R_{[ab]}\neq0$ -- see expression (\ref{Riccit}) next). Finally, by default, the Ricci scalar ($R=g^{ab}R_{ab}$) remains uniquely defined as well.

The relations between the Ricci tensors and the Ricci scalar of the general space and their torsion-free Riemannian associates are obtained directly from (\ref{curvt2}). In particular, after taking successive contractions of the latter, arrive at
\begin{equation}
R_{ab}= \tilde{R}_{ab}+ Q_{ab}+ \tilde{\Gamma}^c{}_{ab}K^d{}_{cd}+ K^c{}_{ab}\tilde{\Gamma}^d{}_{cd}- \tilde{\Gamma}^c{}_{ad}K^d{}_{cb}- K^c{}_{ad}\tilde{\Gamma}^d{}_{cb}  \label{Riccit}
\end{equation}
and
\begin{equation}
R= \tilde{R}+ Q+ g^{ab}\tilde{\Gamma}^c{}_{ab}K^d{}_{cd}+ g^{ab}K^c{}_{ab}\tilde{\Gamma}^d{}_{cd}- g^{ab}\tilde{\Gamma}^c{}_{ad}K^d{}_{cb}- g^{ab}K^c{}_{ad}\tilde{\Gamma}^d{}_{cb}\,.  \label{Riccis}
\end{equation}
The former of the above shows that the symmetric part of the Ricci-Cartan tensor does not necessarily coincide with its Riemannian counterpart (i.e.~$R_{(ab)}\neq\tilde{R}_{ab}$ in general).

In an analogous manner, the successive traces of Eq.~(\ref{Qt}), combined with the definition of the torsion vector (see Eq.~(\ref{torsv}) in \S~\ref{ssTV}), lead to
\begin{equation}
Q_{ab}= \partial_cK^c{}_{ab}- 2\partial_bS_a+ 2K^c{}_{ab}S_c- K^c{}_{ad}K^d{}_{cb}  \label{Qab}
\end{equation}
and
\begin{equation}
Q= g^{ab}\partial_cK^c{}_{ab}- 2g^{ab}\partial_bS_a- 4S^aS_a- K_{abc}K^{cab}\,.  \label{Q}
\end{equation}
Expressions (\ref{Riccit})-(\ref{Q}) reveal that the Ricci tensor and the Ricci scalar of the general space split into a solely Riemannian, an entirely torsional and a mixed part.

\subsection{The Weyl-Cartan tensor}\label{ssW-CT}
When dealing with Riemannian spaces, the curvature (Riemann) tensor decomposes into its trace (described by the Ricci field) and a traceless component that is commonly referred to as the Weyl tensor. In analogy, the Riemann-Cartan curvature tensor splits as~\cite{J}
\begin{equation}
R_{abcd}= C_{abcd}+ R_{a[c}g_{d]b}- R_{b[c}g_{d]a}- {1\over3}\,Rg_{a[c}g_{d]b}\,.  \label{Rsplit}
\end{equation}
The trace-free nature of $C_{abcd}$, which is straightforward to verify, means that the latter may be seen as the Weyl-Cartan curvature tensor in spacetimes with non-vanishing torsion. Note that, by construction (see definition (\ref{Rsplit}) above), $C_{abcd}$ also satisfies the symmetries of the Riemann-Cartan tensor (i.e.~$C_{abcd}=C_{[ab][cd]}$).

\subsection{The Bianchi identities}\label{ssBIs}
When the space has torsion, the generalised Bianchi identities are also known as the Weitzenbock identities and take the form (e.g.~see~\cite{P2})
\begin{equation}
\nabla_{[m}R^{ab}{}_{cd]}= 2R^{ab}{}_{n[c}S^n{}_{dm]} \hspace{10mm} {\rm and} \hspace{10mm} R^a{}_{[bcd]}= -2\nabla_{[d}S^a{}_{bc]}+ 4S^a{}_{m[b}S^m{}_{cd]}\,.  \label{Bianchi}
\end{equation}
Contracting the former of the above twice and using the antisymmetry properties of the torsion and the curvature tensors, we arrive at
\begin{equation}
\nabla^bR_{ba}-{1\over2}\,\nabla_aR= -2S^c{}_{ab}R^b{}_c- S^d{}_{bc}R^{bc}{}_{da}\,.  \label{2cBianchi}
\end{equation}
When the torsion vanishes, this constraint reduces to the familiar conservation law $2\nabla^bR_{ab}-\nabla_aR=0$ of the Riemannian spaces.

\section{Spacetimes with torsion}\label{sS-TT}
If Riemannian spacetimes are the natural hosts of general relativity, their torsional Riemann-Cartan counterparts provide the geometrical framework for the formulation of perhaps the simplest gravitational theory with intrinsic spin. The latter is usually referred to as the Einstein-Cartan, or sometimes as the Einstein-Cartan-Sciama-Kibble, theory.

\subsection{1+3 covariant decomposition}\label{1+3CD}
Let us consider a 4-dimensional spacetime equipped with a Lorentzian metric ($g_{ab}=g_{(ab)}$, with $g_{ab}g^{bc}=\delta_a{}^c$) of signature $(-,+,+,+)$ and introduce a family of observers living along worldlines tangent to the timelike 4-velocity field $u^a$ (normalised so that $u_au^a=-1$). These observers are associated with a symmetric spacelike tensor $h_{ab}=g_{ab}+u_au_b$ (with $h_{ab}u^b=0$, $h_{ab}h^b{}_c=h_{ac}$ and $h_a{}^a=3$). The latter projects orthogonal to the $u_a$-field and essentially defines the metric tensor of the observers' instantaneous 3-dimensional rest-space. On using $u_a$ and $h_{ab}$, one can introduce an irreducible 1+3 splitting of the spacetime into time (along the $u_a$-field) and 3-space (orthogonal to $u_a$). Then, every variable, every operator and every equation can be decomposed into their timelike and spacelike parts (see~\cite{TCM} for a review of the formalism).

The totally antisymmetric Levi-Civita tensor of the 4-D spacetime ($\eta_{abcd}=\eta_{[abcd]}$, with $\eta_{abcd}\eta^{mnpq}=-4!\delta_{[a}{}^m\delta_b{}^n \delta_c{}^p\delta_{d]}{}^q$) splits as
\begin{equation}
\eta_{abcd}=2u_{[a}\varepsilon_{b]cd}- 2\varepsilon_{ab[c}u_{d]}\,,  \label{4DLC}
\end{equation}
with $\varepsilon_{abc}=\eta_{abcd}u^d$ representing the alternating tensor of the 3-D space. Then, it follows that
$\varepsilon_{abc}=\varepsilon_{[abc]}$, that $\varepsilon_{abc}u^c=0$ and that
\begin{equation}
\varepsilon_{abc}\varepsilon^{dmn}= 3!h_{[a}{}^dh_b{}^mh_{c]}{}^n\,.  \label{3DLC}
\end{equation}
Accordingly, $\varepsilon_{abc}\varepsilon^{dmc}= 2h_{[a}{}^dh_{b]}{}^m$, $\varepsilon_{abc}\varepsilon^{dbc}= 2h_a{}^d$ and $\varepsilon_{abc}\varepsilon^{abc}=6$.

\subsection{Temporal and spatial gradients}\label{ssTSGs}
Once a family of observers has been introduced and the spacetime has been split into time and 3-D space, the temporal and spatial derivatives of a general tensor field $T_{ab\cdots}{}^{cd\cdots}=T_{ab\cdots}{}^{cd\cdots}(x^s)$ are defined by
\begin{equation}
\dot{T}_{ab\cdots}{}^{cd\cdots}= u^m\nabla_mT_{ab\cdots}{}^{cd\cdots}  \label{1+3dir1}
\end{equation}
and
\begin{equation}
{\rm D}_mT_{ab\cdots}{}^{cd\cdots}= h_m{}^qh_a{}^fh_b{}^k\cdots h_p{}^ch_r{}^d\cdots\nabla_qT_{fk\cdots}{}^{pr\cdots}\,,  \label{1+3dir2}
\end{equation}
respectively. Note that after applying (\ref{1+3dir2}) to the projection tensor, one can easily show that ${\rm D}_ch_{ab}=0$. In other words, $h_{ab}$ remains invariant under spatial covariant differentiation.

Using the definition of covariant differentiation and the relation between the general connection and the Christoffel symbols (see Eq.~(\ref{Gammas}) in \S~\ref{ssCT}) we can obtain the relations between the temporal and the spatial derivatives in the two spaces. For example, in the case of a covariant second-rank tensor, the time derivatives are related by
\begin{equation}
\dot{T}_{ab}= T^{\prime}_{ab}- u^c\left(K^d{}_{ac}T_{db}+K^d{}_{bc}T_{ad}\right)\,,  \label{tdirs}
\end{equation}
with the primes denoting time-differentiation in terms of the Christoffel symbols of the associated torsion-free space. Similarly, we find that the relation between the spatial derivatives is
\begin{equation}
{\rm D}_cT_{ab}= \tilde{\rm D}_cT_{ab}- h_c{}^fh_a{}^dh_b{}^m \left(K^p{}_{df}T_{pm}+K^p{}_{mf}T_{dp}\right)\,,  \label{sdirs}
\end{equation}
keeping in mind that the ``tildas'' always refer to the associated torsionless space. Note that, when applied to the projection tensor, the former of the above two expressions gives
\begin{equation}
\dot{h}_{ab}= h^{\prime}_{ab}+ 4u^cu^dS_{cd(a}u_{b)}\,,  \label{htdirs}
\end{equation}
where we have also used the symmetries of the contortion tensor (see Eqs.~(\ref{contsym1}) and (\ref{contsym2}) in \S~\ref{ssCT}). Relation (\ref{sdirs}), on the other hand, leads to
\begin{equation}
{\rm D}_ch_{ab}= \tilde{\rm D}_ch_{ab}\,,  \label{hsdirs}
\end{equation}
which guarantees that $\tilde{\rm D}_ch_{ab}=0$ when ${\rm D}_ch_{ab}=0$ and vice-versa. Given that ${\rm D}_ch_{ab}=0$ by construction, we deduce that the projector remains invariant under spatial covariant differentiation both in the general space and in its torsion-free (Riemannian) associate. Then, expression (\ref{3DLC}) guarantees that the 3-D alternating tensor is also covariantly constant (i.e.~${\rm D}_d \varepsilon_{abc}=0=\tilde{\rm D}_d\varepsilon_{abc}$). Finally, we have $\dot{\varepsilon}_{abc}= 3u_{[a}\varepsilon_{bc]d}A^d$, with $A_a=\dot{u}_a$ (see decomposition (\ref{Nbua}) next).

\section{Kinematics with torsion}\label{sKT}
The kinematics of a timelike congruence, as well as that of the associated observers, are monitored through a set of irreducible variables. These describe the individual aspects of the motion and satisfy a set of propagation and constraint equations that are fully geometrical in nature.

\subsection{The irreducible kinematic variables}\label{ssIKVs}
All the information regarding the kinematic of the aforementioned family of observers is encoded in the gradient of their 4-velocity vector. The latter decomposes in to the irreducible components of the motion according to
\begin{equation}
\nabla_bu_a= {\rm D}_bu_a- A_au_b= {1\over3}\,\Theta h_{ab}+ \sigma_{ab}+ \omega_{ab}- A_au_b\,.  \label{Nbua}
\end{equation}
In the above, $\Theta=\nabla^au_a={\rm D}^au_a$ is the volume scalar that monitors the mean separation between the observers' worldlines. In particular, $\Theta$ describes expansion when it takes positive values and contraction in the opposite case. The volume scalar is typically used to introduce a representative length-scale ($a$) along the observers' worldlines, defined by $\dot{a}/a=\Theta/3$. In cosmological studies the latter is known as the scale factor and it is directly related to the Hubble parameter (i.e.~$\dot{a}/a=H$). The symmetric and trace-free shear tensor $\sigma_{ab}={\rm D}_{\langle b}u_{a\rangle}$, which is spacelike by construction (i.e.~$\sigma_{ab}u^b=0$), reflects kinematic anisotropies. When applied to a fluid element, in particular, the shear describes changes in its shape under constant volume. The antisymmetric vorticity tensor $\omega_{ab}={\rm D}_{[b}u_{a]}$ is also spacelike (i.e.~$\omega_{ab}u^b=0$) and monitors the rotational behaviour of the observers' worldlines. Moreover, the associated vorticity vector $\omega_a=\varepsilon_{abc}\omega^{bc}/2$ (with $\omega_au^a=0$, since $\varepsilon_{abc}u^c=0$) defines the direction of the rotational axis. Finally, $A_a=u^b\nabla_bu_a$ is the 4-acceleration vector, with $A_au^a=0$ as well. Also, following \S~\ref{ssAGCs}, the 4-acceleration vanishes when the observers' worldlines are autoparallel curves.

\subsection{Cartan vs Riemannian variables}\label{ssCvsRVs}
Confining to the Riemannian (torsion-free) associate of our general space, expression (\ref{Nbua}) takes the form
\begin{equation}
\tilde{\nabla}_bu_a= \tilde{\rm D}_bu_a- \tilde{A}_au_b= {1\over3}\,\tilde{\Theta} h_{ab}+ \tilde{\sigma}_{ab}+ \tilde{\omega}_{ab}- \tilde{A}_au_b\,,  \label{tNbua}
\end{equation}
where the tilded variables are defined in a way exactly analogous to that of their non-tilded counterparts. Also, by construction we have $\tilde{\sigma}_{ab}u^b=0= \tilde{\omega}_{ab}u^b=\tilde{A}_au^a$. Using the symmetries of the contortion tensor (see Eqs.~(\ref{contsym1}) and (\ref{contsym2}) in \S~\ref{ssCT}), the relations between the two sets of variables given in Eqs.~(\ref{Nbua}) and (\ref{tNbua}) are\footnote{Analogous relations, between the purely Riemannian and the torsional kinematic variables, have been also obtained in~\cite{MT}, though the conventions used there by the authors are generally different from those adopted here.}
\begin{equation}
\Theta= \tilde{\Theta}+ 2S_au^a\,, \hspace{10mm} \sigma_{ab}= \tilde{\sigma}_{ab}- 2h_{(a}{}^ch_{b)}{}^dS_{cdm}u^m- {2\over3}\,S_cu^ch_{ab}\,, \label{kin1}
\end{equation}
\begin{equation}
\omega_{ab}= \tilde{\omega}_{ab}- h_{[a}{}^ch_{b]}{}^dS_{mcd}u^m \hspace{10mm} {\rm and} \hspace{10mm} A_a= \tilde{A}_a+ 2S_{(bc)a}u^bu^c\,.  \label{kin2}
\end{equation}
According to these relations, a fully antisymmetric torsion tensor means that $\Theta=\tilde{\Theta}$, $\sigma_{ab}=\tilde{\sigma}_{ab}$, $A_a=\tilde{A}_a$ (since $S_a=0=S_{(ab)c}$ when $S_{abc}=S_{[abc]}$) and only $\omega_{ab}\neq\tilde{\omega}_{ab}$. More specifically, following Eq.~(\ref{kin1}a), the two volume scalars coincide when the torsion vector vanishes (i.e.~when $S^a{}_{bc}$ is traceless -- see \S~\ref{ssTV} earlier), or when $S_a$ is nonzero but spacelike (i.e.~for $S_au^a=0$). In general, however, $\Theta\neq\tilde{\Theta}$ and the same is also true for the shear, the vorticity and the 4-acceleration (i.e.~$\sigma_{ab}\neq\tilde{\sigma}_{ab}$, $\omega_{ab}\neq\tilde{\omega}_{ab}$ and $A_a\neq\tilde{A}_a$). It is also worth pointing out that (\ref{kin1}b) and (\ref{kin2}) ensure that $\tilde{\sigma}_{ab}u^b=0= \tilde{\omega}_{ab}u^b=\tilde{A}_au^a$, thus guaranteeing that $\tilde{\sigma}_{ab}$, $\tilde{\omega}_{ab}$ and $\tilde{A}_a$ are spacelike quantities as well. Finally, we note that expression (\ref{kin2}b) is consistent with relation (\ref{htdirs}), between the time derivatives of the projector, while it provides the relation $\dot{\varepsilon}_{abc}=\varepsilon^{\prime}_{abc}+ 6u_{[a}\varepsilon_{bc]d}S_{(sf)}{}^du^su^f$ between the temporal derivatives of the spatial Levi-Civita tensor.\footnote{Throughout this work we assume a non-tilted spacetime. A Lorentz boost of the observers 4-velocity will also affect the irreducible variables of the motion. When the relative velocity between the two frames is not relativistic, the changes (generally) resemble those seen in Eqs.~(\ref{kin1}) and (\ref{kin2}) above (e.g.~see Appendix A2 in~\cite{TCM}).}

\section{Kinematic evolution}\label{sKE}
With the exception of the 4-acceleration, the time evolution of the kinematic variables defined in the previous section is obtained after applying the Ricci identity to the observers' 4-velocity vector. In particular, the timelike component of the resulting expression leads to the propagation formulae of $\Theta$, $\sigma_{ab}$ and $\omega_{ab}$, while its spacelike part provides the associated constraints.

\subsection{The timelike Ricci identities}\label{ssTRIs}
In spaces that allow for nonzero torsion, the Ricci identity takes the form (e.g.~see~\cite{Pa})\footnote{In a general (not necessarily metric) space with asymmetric connection $\Gamma^a{}_{bc}$, applying the Ricci identity to an arbitrary contravariant vector $u^a$ leads to the expression $2\nabla_{[c}\nabla_{b]}u^a=-R^a{}_{dbc}u^d- 2S^d{}_{bc}\nabla_du^a$. When a metric is introduced into the space (with $\nabla_cg_{ab}=0$), the above relation combines with (\ref{tRicci}) to give $R_{abcd}=-R_{bacd}$.}
\begin{equation}
2\nabla_{[c}\nabla_{b]}u_a= R^d{}_{abc}u_d- 2S^d{}_{bc}\nabla_du_a\,,  \label{tRicci}
\end{equation}
which in the absence of torsion reduces to the more familiar Riemannian expression $2\tilde{\nabla}_{[a} \tilde{\nabla}_{b]}u_c=\tilde{R}_{abcd}u^d$ (given the increased symmetries of the corresponding curvature tensor). Contracting Eq.~(\ref{tRicci}) along $u_c$, using decomposition (\ref{Nbua}) and employing some fairly straightforward algebra, leads to the intermediate relation\footnote{When deriving expression (\ref{aux1}), one also needs to use the following auxiliary relation
\begin{equation}
\nabla_bA_a= {\rm D}_bA_a+ {1\over3}\,\Theta u_aA_b+ u_a(\sigma_{bc}-\omega_{bc})A^c- (A_au_b)^{\cdot}+ A_aA_b\,,  \label{aux3}
\end{equation}
between the gradients of the 4-acceleration vector.}
\begin{eqnarray}
\left(\nabla_bu_a\right)^{\cdot}&=& -{1\over9}\,\Theta^2h_{ab}- R_{cadb}u^cu^d- {2\over3}\,\Theta(\sigma_{ab}+\omega_{ab})- \sigma_{ca}\sigma_b{}^c- \omega_{ca}\omega_b{}^c+ 2\sigma_{c[a}\omega_{b]}{}^c \nonumber\\ &&+{\rm D}_bA_a+ {2\over3}\,\Theta u_{\langle a}A_{b\rangle}+ 2u_{\langle a}\sigma_{b\rangle c}A^c- 2u_{[a}\omega_{b]c}A^c- \left(A_au_b\right)^{\cdot}+ A_aA_b\nonumber\\ &&-{2\over3}\,\Theta S_{abc}u^c+ 2\left({1\over3}\,\Theta u_a-A_a\right) u^cu^dS_{cdb}+2(\sigma_a{}^c+\omega_a{}^c)u^dS_{cdb}\,.  \label{aux1}
\end{eqnarray}
Substituting into the left-hand side of the above the decomposition of the 4-velocity gradient (into the irreducible kinematic variables -- see Eq.~(\ref{Nbua}) in \S~\ref{ssIKVs}) and keeping in mind that $\dot{h}_{ab}=2u_{\langle a}A_{b\rangle}$, gives
\begin{eqnarray}
{1\over3}\,\dot{\Theta}h_{ab}+ \dot{\sigma}_{ab}+ \dot{\omega}_{ab}&=& -{1\over9}\,\Theta^2h_{ab}- R_{cadb}u^cu^d- {2\over3}\,\Theta(\sigma_{ab}+\omega_{ab})- \sigma_{ca}\sigma_b{}^c- \omega_{ca}\omega_b{}^c+ 2\sigma_{c[a}\omega_{b]}{}^c\nonumber\\ &&+{\rm D}_bA_a+ 2u_{\langle a}\sigma_{b\rangle c}A^c- 2u_{[a}\omega_{b]c}A^c+ A_aA_b\nonumber\\ &&-{2\over3}\,\Theta S_{abc}u^c+ 2\left({1\over3}\,\Theta u_a-A_a\right) u^cu^dS_{cdb}+2(\sigma_a{}^c+\omega_a{}^c)u^dS_{cdb}\,.  \label{aux2}
\end{eqnarray}
Finally, projecting the latter orthogonal to the $u_a$-field and using the symmetries of the curvature tensor (see \S~\ref{ssR-CT} earlier), we arrive at
\begin{eqnarray}
{1\over3}\,\dot{\Theta}h_{ab}+ h_{\langle a}{}^ch_{b\rangle}{}^d\dot{\sigma}_{cd}+ h_{[a}{}^ch_{b]}{}^d\dot{\omega}_{cd}&=& -{1\over9}\,\Theta^2h_{ab}- R_{acbd}u^cu^d- {2\over3}\,\Theta(\sigma_{ab}+\omega_{ab})\nonumber\\ &&-\sigma_{ca}\sigma_b{}^c- \omega_{ca}\omega_b{}^c+ 2\sigma_{c[a}\omega_{b]}{}^c+ {\rm D}_bA_a+ A_aA_b\nonumber\\ &&-{2\over3}\,\Theta h_a{}^ch_b{}^dS_{cdm}u^m- 2A_au^cu^dh_b{}^mS_{cdm}\nonumber\\ &&+2(\sigma_a{}^c+\omega_a{}^c)u^dh_b{}^mS_{cdm}\,.  \label{kinev}
\end{eqnarray}
given that $h_a{}^ch_b{}^d\dot{\sigma}_{cd}=h_{\langle a}{}^ch_{b\rangle}{}^d\dot{\sigma}_{cd}$ and $h_a{}^ch_b{}^d\dot{\omega}_{cd}=h_{[a}{}^ch_{b]}{}^d\dot{\omega}_{cd}$. Note that the terms in the first two lines on the right-hand side have direct Riemannian analogues, whereas those in the last two lines are explicitly due to the presence of torsion. Also, the curvature tensor contains a entirely Riemannian, an exclusively torsional and a mixed component (see Eq.~(\ref{curvt2}) in~\S~\ref{ssCT} earlier).

Expression (\ref{kinev}) governs the full kinematic evolution of observers living in spacetimes with nonzero torsion, with no prior assumptions regarding the nature of the torsion tensor (or its coupling to the spin of the matter). As we will show next, the trace, the projected symmetric trace-free and the projected antisymmetric components of (\ref{kinev}) provide the evolution formulae of the volume scalar ($\Theta$), of the shear tensor ($\sigma_{ab}$) and of the vorticity tensor ($\omega_{ab}$) respectively.

\subsection{The Raychaudhuri equation}\label{ssRE}
Taking the trace of Eq.~(\ref{kinev}), while recalling that $R_{ab}=R^c{}_{acb}$, that $S_{abc}=S_{a[bc]}$ and that $S^b{}_{ba}=-S_a$, we obtain the expression
\begin{eqnarray}
\dot{\Theta}&=& -{1\over3}\,\Theta^2- R_{(ab)}u^au^b- 2\left(\sigma^2-\omega^2\right)+ {\rm D}_aA^a+ A_aA^a\nonumber\\ &&+{2\over3}\,\Theta S_au^a- 2S_{(ab)c}u^au^bA^c- 2S_{\langle ab\rangle c}\sigma^{ab}u^c+ 2S_{[ab]c}\omega^{ab}u^c\,,  \label{Ray1}
\end{eqnarray}
which is the analogue of the Raychaudhuri equation in spaces with nonzero torsion. Note that $\sigma^2=\sigma_{ab}\sigma^{ab}/2$ and $\omega^2=\omega_{ab}\omega^{ab}/2=\omega_a\omega^a$ by definition. Also, only the symmetric part of the Ricci tensor (which is nevertheless torsion-dependent -- see Eq.~(\ref{Riccit}) earlier) contributes to Raychaudhuri's formula. Finally, we should point out that the terms in the first line on the right-hand side of the above have Riemannian analogues (e.g.~see \S~1.3.1 in~\cite{TCM}), while those in the second line are explicitly due to torsion.

The Raychaudhuri equation is the key formula of gravitational contraction/expansion and it has played a fundamental role in the formulation of the various singularity theorems (e.g.~see~\cite{HE}). Following (\ref{Ray1}), positive terms on its right-hand side inhibit the contraction, or assist the expansion. Negative terms, on the other hand, act in the opposite way. With the exception of~\cite{KS}, the torsional analogue of Raychaudhuri's formula has been derived after imposing certain symmetry conditions, namely spatial homogeneity and isotropy (FRW models), spatial homogeneity (Bianchi-type models)~\cite{K1,K2}, or for the case of the Weyssehoff fluid~\cite{SH,BHL}. Also, typically, the field equations are split into a purely Riemannian and a torsion/spin part. Here, as yet, we have not made any assumptions of this kind. Expression (\ref{Ray1}) applies to a general Riemann-Cartan spacetime, with no a priori restrictions imposed on the nature of the gravitational interaction or on the relation between torsion and spin.\footnote{An alternative form of Raychaudhuri's formula with a general torsion field was recently given in~\cite{hL}.} Once these are specified, it will be possible to decode the effects of torsion in more detail (see \S~\ref{ssE-CKs} and \S~\ref{ssREE-CUs} below).

Raychaudhuri's formula can simplify considerably under certain symmetry conditions. For instance, in the special case of a totally antisymmetric torsion tensor, Eq.~(\ref{Ray1}) reads
\begin{equation}
\dot{\Theta}= -{1\over3}\,\Theta^2- R_{(ab)}u^au^b- 2\left(\sigma^2-\omega^2\right)+ {\rm D}_aA^a+ A_aA^a+ 2S_{abc}\omega^{ab}u^c\,,  \label{Ray2}
\end{equation}
given that $S_a=0=S_{(ab)c}=S_{\langle ab\rangle c}$ when $S_{abc}=S_{[abc]}$. Also, assuming that the worldlines tangent to the $u_a$-field are autoparallel curves, the 4-acceleration vanishes identically (i.e.~$A_a=0$). In addition, when the aforementioned autoparallel congruence is also shear-free and irrotational, we may set $\sigma_{ab}=0= \omega_{ab}$ as well. Then, for standard torsion (with $S^a{}_{bc}=S^a{}_{[bc]}$ and $S_a\neq0$), expression (\ref{Ray1}) reduces to
\begin{equation}
\dot{\Theta}= -{1\over3}\,\Theta^2- R_{(ab)}u^au^b+ {2\over3}\,\Theta S_au^a\,,  \label{Ray3}
\end{equation}
with only the last term having explicit torsional nature. Therefore, when the inner product $S_au^a$ takes positive values, it tends to speed up the contraction/expansion of a self-gravitating medium. In the opposite case the effect is reversed, while for purely spacelike torsion vectors this term vanishes identically. By construction, the above expression also monitors the expansion/contraction rate of spatially homogeneous and isotropic spacetimes, which may be seen as the torsional analogues of the familiar FRW universes. In that case the torsion vector has to be purely timelike, since otherwise its presence would have destroyed the isotropy of the model's spatial hypersurfaces. According to (\ref{Ray3}), when $2\Theta S_au^a-3R_{(ab)}u^au^b>0$, worldline focusing and the initial singularity can be averted.

We finally note that when $S_a$ vanishes, namely when $S^a{}_{bc}$ is trace-free -- see \S~\ref{ssTV} earlier, there are no explicit torsion terms on the right-hand side of the above. Then, the effects of spacetime torsion come solely from the non-Riemannian components of the the Ricci-Cartan tensor (see Eqs.~(\ref{Riccit}), (\ref{Qab}) in \S~\ref{ssRicci-CT}) and those of the volume scalar (see Eq.~(\ref{kin1}a) in \S~\ref{ssCvsRVs}). This also true when dealing with a spacelike torsion vector (i.e.~for $S_au^a=0$).

\subsection{Shear and vorticity evolution}\label{ssSVE}
The symmetric trace-free and the antisymmetric parts of the general expression (\ref{kinev}), provide the respective evolution formulae of the shear and the vorticity tensors in spacetimes with nonzero torsion. In particular we obtain
\begin{eqnarray}
h_{\langle a}{}^ch_{b\rangle}{}^d\dot{\sigma}_{cd}&=& - {2\over3}\,\Theta\sigma_{ab}- \sigma_{c\langle a}\sigma_{b\rangle}{}^c- \omega_{c\langle a}\omega_{b\rangle}{}^c+ {\rm D}_{\langle b}A_{a\rangle}+ A_{\langle a}A_{b\rangle}- R_{\langle a}{}^c{}_{b\rangle}{}^d u_cu_d\nonumber\\ &&-{2\over3}\,\Theta h_{\langle a}{}^ch_{b\rangle}{}^dS_{cdm}u^m- 2A_{\langle a} u^cu^dh_{b\rangle}{}^mS_{cdm}\nonumber\\ &&+2\left(\sigma_{\langle a}{}^c+\omega_{\langle a}{}^c\right) u^dh_{b\rangle}{}^mS_{cdm}\,,  \label{shev1}
\end{eqnarray}
for the shear and
\begin{eqnarray}
h_{[a}{}^ch_{b]}{}^d\dot{\omega}_{cd}&=& -{2\over3}\,\Theta\omega_{ab}+ 2\sigma_{c[a}\omega_{b]}{}^c+ {\rm D}_{[b}A_{a]}-R_{[a}{}^c{}_{b]}{}^d u_cu_d- {2\over3}\,\Theta h_{[a}{}^ch_{b]}{}^dS_{cdm}u^m\nonumber\\ &&-2A_{[a}u^cu^dh_{b]}{}^mS_{cdm}+ 2\left(\sigma_{[a}{}^c+\omega_{[a}{}^c\right) u^dh_{b]}{}^mS_{cdm}\,,  \label{vortev1}
\end{eqnarray}
for the vorticity. The former of these expressions monitors distortions in the shape of the $u_a$-congruence, which occur under constant volume, while the latter governs the rotational behaviour of these worldlines. As with the Raychaudhuri equation before, when dealing with autoparallel curves, all the 4-acceleration terms on right-hand sides of (\ref{shev1}) and (\ref{vortev1}) vanish identically. Also note that, in both of the above, the curvature tensor is given by Eq.~(\ref{curvt2}).

We may analyse the curvature terms on the right-hand side of Eqs.~(\ref{shev1}) and (\ref{vortev1}) further, by employing the decomposition of the Riemann-Cartan curvature tensor into its Weyl and Ricci parts (see expression (\ref{Rsplit}) in \S~\ref{ssW-CT} earlier). In particular, keeping in mind the traceless nature of the Weyl tensor, we arrive at
\begin{eqnarray}
h_{\langle a}{}^ch_{b\rangle}{}^d\dot{\sigma}_{cd}&=& - {2\over3}\,\Theta\sigma_{ab}- \sigma_{c\langle a}\sigma_{b\rangle}{}^c- \omega_{c\langle a}\omega_{b\rangle}{}^c+ {\rm D}_{\langle b}A_{a\rangle}+ A_{\langle a}A_{b\rangle}+ {1\over2}\,h_{\langle a}{}^ch_{b\rangle}{}^dR_{cd}\nonumber\\ &&-C_{\langle a}{}^c{}_{b\rangle}{}^d u_cu_d- {2\over3}\,\Theta h_{\langle a}{}^ch_{b\rangle}{}^dS_{cdm}u^m- 2A_{\langle a} u^cu^dh_{b\rangle}{}^mS_{cdm}\nonumber\\ &&+2\left(\sigma_{\langle a}{}^c+\omega_{\langle a}{}^c\right) u^dh_{b\rangle}{}^mS_{cdm}\,,  \label{shev2}
\end{eqnarray}
and
\begin{eqnarray}
h_{[a}{}^ch_{b]}{}^d\dot{\omega}_{cd}&=& -{2\over3}\,\Theta\omega_{ab}+ 2\sigma_{c[a}\omega_{b]}{}^c+ {\rm D}_{[b}A_{a]}+ {1\over2}\,h_{[a}{}^ch_{b]}{}^dR_{cd}- C_{[a}{}^c{}_{b]}{}^d u_cu_d\nonumber\\ &&-{2\over3}\,\Theta h_{[a}{}^ch_{b]}{}^dS_{cdm}u^m- 2A_{[a}u^cu^dh_{b]}{}^mS_{cdm}\nonumber\\ &&+2\left(\sigma_{[a}{}^c+\omega_{[a}{}^c\right) u^dh_{b]}{}^mS_{cdm}\,,  \label{vortev2}
\end{eqnarray}
respectively. An immediate conclusion following from expressions (\ref{shev1})-(\ref{vortev2}) is that spacetime torsion can source both shear and rotational anisotropies, which is not surprising. We also note that he symmetric and trace-free tensor $E_{\langle ab\rangle}=C_{\langle a}{}^c{}_{b\rangle}{}^d u_cu_d$ seen in Eq.~(\ref{shev2}) may be interpreted as the electric component of the Weyl tensor in spacetimes with non-vanishing torsion. Then, the symmetry properties $C_{abcd}=C_{[ab][cd]}$ ensure that $E_{\langle ab\rangle}u^a=0$. On the other hand, the fact that $C_{abcd}\neq C_{cdab}$ implies that $E_{[ab]}=C_{[a}{}^c{}_{b]}{}^d u_cu_d\neq0$, in contrast to its Riemannian analogue. We also point out that the second and third-line terms on the right-hand side of Eq.~(\ref{shev2}) are explicitly due to the presence of spacetime torsion, while the rest have Riemannian analogues. The difference with (\ref{vortev2}) is that there the Ricci and the Weyl terms are also purely torsional with no Riemannian analogues (e.g.~compare to Eqs~(1.3.4) and (1.3.5) in \S~1.3.1 of~\cite{TCM}).

\section{Kinematic constraints}\label{sKCs}
The three evolution formulae of the previous section are supplemented by an equal number of constraints. These hold on the observer's 3-D rest-space and they are obtained after applying the Ricci identity to the 4-velocity vector and taking the spacelike part of the resulting expression.

\subsection{The spacelike Ricci identities}\label{ssSRIs}
Contracting Eq.~(\ref{tRicci}) with the 3-D Levi-Civita tensor gives
\begin{equation}
\varepsilon_{cda}\nabla^c\nabla^du_b= -{1\over2}\,\varepsilon_{cda}R_{mb}{}^{cd}u^m+ \varepsilon_{cda}S^{mcd}\nabla_mu_b\,.  \label{aux4}
\end{equation}
Substituting decomposition (\ref{Nbua}) into the above, projecting the resulting expression orthogonally to the $u_a$-field and keeping in mind that $\omega_{ab}= \varepsilon_{abc}\omega^c$, leads to the intermediate relation
\begin{eqnarray}
{1\over3}\,\varepsilon_{abc}{\rm D}^c\Theta- \varepsilon_{cda}{\rm D}^c\sigma_b{}^d+ ({\rm D}^c\omega_c)h_{ab}- {\rm D}_b\omega_a&=& 2\omega_aA_b- {1\over2}\,\varepsilon_{cda}R_{bm}{}^{cd}u^m- {1\over3}\,\Theta\,\varepsilon_{cda}h_{bm}S^{mcd}\nonumber\\ &&-\varepsilon_{cda}\sigma_{bm}S^{mcd}+ \varepsilon_{cda}\varepsilon_{bmn}\omega^{[m}S^{n]cd} \nonumber\\ &&+\varepsilon_{cda}A_bu_mS^{mcd}\,,  \label{aux5}
\end{eqnarray}
where only the left-hand-side terms and the first two on the right-hand side have Riemannian analogues. The above provides the general constraint equation obeyed by the spatial gradients of the kinematic variables on the observers' instantaneous 3-D rest-space. Taking the trace, the antisymmetric, as well as the symmetric and trace-free component of (\ref{aux5}) leads to a scalar, a vector and a (traceless) tensor constraint respectively.

\subsection{The scalar constraint}\label{ssSC}
Isolating the trace of expression (\ref{aux5}), taking into account the properties of the 3-D Levi-Civita tensor (recall that $\varepsilon_{abc}=\varepsilon_{[abc]}$ and $\varepsilon_{abc}u^c=0$ -- see \S~\ref{1+3CD} before), using the definition of the torsion vector (see Eq.~(\ref{torsv}) in \S~\ref{ssTV}) and decomposition (\ref{Rsplit}), leads to the scalar constraint
\begin{eqnarray}
{\rm D}_a\omega^a&=& A_a\omega^a+ {1\over4}\,\varepsilon_{abc}u_dC^{d[abc]}- {1\over6}\,\Theta\varepsilon_{abc}S^{[abc]}- {1\over2}\,\varepsilon_{abc}\sigma^{d[a}S_d{}^{bc]}+ S_a\omega^a- S_{(ab)c}u^au^b\omega^c\nonumber\\ &&+{1\over2}\,\varepsilon_{abc}u^dA^{[a}S_d{}^{bc]}\,,  \label{con1}
\end{eqnarray}
which determines the 3-divergence of the vorticity vector in the presence of torsion. Note that, in addition to the torsion terms, the Weyl-curvature term also vanish in Riemannian spaces (since $C_{a[bcd]}=0$ there).

\subsection{The vector constraint}\label{ssVC}
Taking the antisymmetric component of Eq.~(\ref{aux5}), using relation (\ref{3DLC}) and decomposition (\ref{Rsplit}), while keeping in mind that ${\rm D}_ch_{ab}=0$, the traceless nature of the Weyl tensor and setting ${\rm curl}v_a= \varepsilon_{abc}{\rm D}^bv^c$ (for any spacelike vector $v_a$), provides the vector constraint
\begin{eqnarray}
{2\over3}\,{\rm D}_a\Theta&=& {\rm D}^b\sigma_{ab}- {\rm curl}\omega_a- 2\varepsilon_{abc}A^b\omega^c- h_a{}^bR_{cb}u^c+ {2\over3}\,\Theta h_a{}^{[c}h_b{}^{d]}S^b{}_{cd}+ 2h_a{}^{[c}\sigma_b{}^{d]}S^b{}_{cd}\nonumber\\ &&-2h_a{}^{[c}\varepsilon_b{}^{d]m}S^b{}_{cd}\omega_m- 2h_a{}^{[c}A^{d]}u_bS^b{}_{cd}\,,  \label{con2}
\end{eqnarray}
obeyed by the spatial gradient of the volume scalar in spacetimes with nonzero torsion. Here, only the torsion terms have non-Riemannian analogues (e.g.~see \S~1.3.1 in~\cite{TCM}).

\subsection{The tensor constraint}\label{ssTC}
Finally, after taking the symmetric and trace-free part of expression (\ref{aux5}) and setting ${\rm curl}v_{ab}=\varepsilon_{cd\langle a}{\rm D}^cv^d{}_{b\rangle}$ for any spacelike tensor $v_{ab}$, we arrive at the (traceless) tensor constraint
\begin{eqnarray}
{\rm curl}\sigma_{ab}&=& -{\rm D}_{\langle b}\omega_{a\rangle}- 2A_{\langle a}\omega_{b\rangle}+ {1\over2}\,\varepsilon_{cd\langle a}C_{b\rangle m}{}^{[cd]}u^m+ {1\over3}\,\Theta\varepsilon_{cd\langle a}h_{b\rangle m} S^{mcd}+ \varepsilon_{cd\langle a}\sigma_{b\rangle m} S^{mcd}\nonumber\\ &&-\varepsilon_{cd\langle a}\varepsilon_{b\rangle mn}\omega^{[n} S^{m]cd}- \varepsilon_{cd\langle a} A_{b\rangle}u_mS^{mcd}\,.  \label{con3}
\end{eqnarray}
which determines the curl of the shear in spacetimes with non-vanishing torsion. Note that the symmetric and trace-free tensor $H_{\langle ab\rangle}= \varepsilon_{cd\langle a}C_{b\rangle m}{}^{cd}u^m/2$ may be seen as the magnetic component of the Weyl tensor in spacetimes with nonzero torsion and reduces to its standard Riemannian counterpart in a torsion-free environment. In addition, by construction we have $H_{\langle ab\rangle}u^b=0$. Finally, as in Eq.~(\ref{con2}) previously, only the torsion terms on the right-hand side of the above have no Riemannian analogues (e.g.~see \S~1.3.1 in~\cite{TCM}).

Before closing this section, we should emphasise that, so far, our study and our results have been purely geometrical in nature. We have analysed the kinematics of timelike worldlines in spacetimes with nonzero torsion, and derived the associated evolution and constraint equations, without making any prior assumptions neither about the material content of our spacetime, nor about the nature of the interaction between the matter and the geometry of the host space. Once the field equations and the material content of the spacetime have been specified, our formulae can be used to describe the kinematics of the associated Einstein-Cartan universe. Also note that, after employing relation (\ref{Gammas}) and the equations given in \S~\ref{ssAGCs}, \S~\ref{ssRicci-CT}, \S~\ref{ssTSGs} and \S~\ref{ssCvsRVs}, one can in principle separate the purely Riemannian from the explicitly torsional part of our kinematic formulae. Finally, in the absence of torsion, the full symmetries of the Riemann and the Weyl tensor are restored. Then, expressions (\ref{Ray1})-(\ref{vortev2}) and (\ref{con1})-(\ref{con3}) reduce to their standard Riemannian counterparts (see~\S~1.3.1 in~\cite{TCM} for a direct comparison).

\section{Einstein-Cartan universes}\label{sE-CUs}
The Einstein-Cartan gravity, or the Einstein-Cartan-Kibble-Sciama theory, as it is also referred to, is probably the simplest extension of general relativity that also accounts for the spin of the matter. As noted in the introduction, it is a viable theory that is expected to depart significantly from Einstein's gravity for matter densities well above the nuclear threshold.

\subsection{The Einstein-Cartan field 
equations}\label{ssE-CFEs}
In the Einstein-Cartan theory we deal with a set of two field equations: one relating the curvature of the spacetime to the energy density of the material component and another coupling the spacetime torsion to the matter spin. The former maintains the form of its general relativistic counterpart, but without the a priori symmetry of the Ricci and the energy-momentum tensors. In particular, for zero cosmological constant, the Einstein-Cartan field equations read~\cite{P1}
\begin{equation}
R_{ab}- {1\over2}\,Rg_{ab}= \kappa T_{ab}\,,  \label{ECFE1}
\end{equation}
where $R_{ab}$ and $R$ are given by (\ref{Riccit}) and (\ref{Riccis}) respectively, while $\kappa=8\pi G$ and $T_{ab}$ is the canonical energy-momentum tensor of the matter. Going back to expression (\ref{ECFE1}) we find that $R=-\kappa T$. Then, the Einstein-Cartan field equations recast as
\begin{equation}
R_{ab}= \kappa T_{ab}- {1\over2}\,\kappa Tg_{ab}\,.  \label{ECFE2}
\end{equation}

The canonical spin tensor ($s_{abc}$) and the associated spin vector ($s_a$) of the matter relate with their corresponding torsion tensor and vector though the Cartan field equations, namely~\cite{P1}
\begin{equation}
S_{abc}- S_bg_{ca}+ S_cg_{ab}= -{1\over2}\,\kappa s_{bca} \hspace{10mm} {\rm and} \hspace{10mm} S_a= -{1\over4}\,\kappa s_a\,.  \label{CFE1}
\end{equation}
Recall that $S_a=S^b{}_{ab}=-S^b{}_{ba}$ (see \S~\ref{ssTV} earlier). Also, $s_{abc}=s_{[ab]c}$ by construction and $s_a=s_{ba}{}^b=-s_{ab}{}^b$ defines the canonical spin vector. On using the latter of the above expressions, the Cartan field equations (i.e.~expression~(\ref{CFE1}a)) assume the alternative form\footnote{In~\cite{P1}, as well in several other papers working on Einstein-Cartan gravity, the metric-signature convention is ($+,-,-,-$). The transformation rules between the two signatures for the key tensors and operators are:
\begin{eqnarray}
&g_{ab}\rightarrow-g_{ab}\,, \hspace{10mm} u_a\rightarrow-u_a\,, \hspace{10mm} h_{ab}\rightarrow-h_{ab}\,, \hspace{10mm} \eta_{abcd}\rightarrow\eta_{abcd}\,,&  \label{signtrans1} \\ &\partial_a\rightarrow\partial_a\,, \hspace{10mm} \nabla_a\rightarrow\nabla_a\,,  \hspace{10mm} S^a{}_{bc}\rightarrow S^a{}_{bc}\,,&  \label{signtrans2} \\ &R_{abcd}\rightarrow-R_{abcd}\,, \hspace{10mm} T_{ab}\rightarrow T_{ab}\,, \hspace{10mm} s_{ab}{}^c\rightarrow s_{ab}{}^c\,.&  \label{signtrans3}
\end{eqnarray}
Note that the transformations of the metric tensors (see (\ref{signtrans1}a) and (\ref{signtrans1}c)) ensure that raising an lowering indices changes the sign of the quantities involved (e.g.~$R_{ab}\rightarrow R_{ab}$, $R\rightarrow-R$, $S_{abc}\rightarrow-S_{abc}$, etc. -- see also~\cite{BHL}).}
\begin{equation}
S_{abc}= -{1\over4}\,\kappa\,(2s_{bca}+g_{ca}s_b-g_{ab}s_c)\,.  \label{CFE2}
\end{equation}
Note that, in line with Eq.~(\ref{CFE1}a), a vanishing torsion vector implies that $S_{abc}=-\kappa s_{bca}/2$, which guarantees that the spin vector also vanishes. Moreover, when dealing with a totally antisymmetric torsion tensor (with $S_a=0$ as a result), we have $S_{abc}=-\kappa s_{abc}/2$ to ensure the total antisymmetry of the spin tensor as well.

\subsection{Einstein-Cartan kinematics}\label{ssE-CKs}
Starting from the Einstein-Cartan field equations it is straightforward to arrive at the following algebraic relations between the Ricci and the stress-energy tensors
\begin{eqnarray}
&R_{(ab)}u^au^b= \kappa T_{(ab)}u^au^b+ {1\over2}\,\kappa T\,, \hspace{10mm} \hspace{10mm} h_a{}^bR_{bc}u^c= \kappa h_a{}^bT_{bc}u^c\,,&  \label{algECFE1} \\ &h_{\langle a}{}^ch_{b\rangle}{}^dR_{cd}= \kappa h_{\langle a}{}^ch_{b\rangle}{}^dT_{cd} \hspace{10mm} {\rm and} \hspace{10mm} h_{[a}{}^ch_{b]}{}^dR_{cd}= \kappa h_{[a}{}^ch_{b]}{}^dT_{cd}\,.&  \label{algECFE2}
\end{eqnarray}
Similarly, the Cartan field equations lead to auxiliary relations between torsion and spin. For example, employing (\ref{CFE2}), the scalars $S_au^a$, $S_{(ab)c}u^au^bA^c$, $S_{(ab)c}\sigma^{ab}u^c$ and $S_{[ab]c}\omega^{ab}u^c$ found on the right-hand side of Raychaudhuri's formula (see Eq.~(\ref{Ray1}) in \S~\ref{ssRE}) can be replaced by
\begin{eqnarray}
&S_au^a= -{1\over4}\,\kappa\,s_au^a\,, \hspace{10mm} S_{(ab)c}u^au^bA^c= {1\over4}\,\kappa\, (2s_{a(bc)}A^au^bu^c-s_aA^a)\,,&  \label{algCFE1} \\
&S_{\langle ab\rangle c}\sigma^{ab}u^c= {1\over2}\,\kappa\, s_{a\langle bc\rangle}u^a\sigma^{bc} \hspace{10mm} {\rm and} \hspace{10mm} S_{[ab]c}\omega^{ab}u^c= -{1\over2}\,\kappa\, s_{a[bc]}u^a\omega^{bc}\,.&  \label{algCFE2}
\end{eqnarray}
This way one can replace the torsion terms in all the kinematic formulae given in \S~\ref{sKE} and \S~\ref{sKCs} with spin-related variables. Overall, using the Einstein-Cartan and the Cartan field equations, all the geometrical (i.e.~the curvature and the torsion) quantities are replaced with matter variables.

Substituting (\ref{algECFE1}a), together with the auxiliary relations (\ref{algCFE1}) and (\ref{algCFE2}), into the right-hand side of (\ref{Ray1}) leads to the Raychaudhuri equation of an Einstein-Cartan universe, namely
\begin{eqnarray}
\dot{\Theta}&=& -{1\over3}\,\Theta^2- \kappa T_{(ab)}u^au^b- {1\over2}\,\kappa T- 2\left(\sigma^2-\omega^2\right)+ {\rm D}_aA^a+ A_aA^a \nonumber\\ &&-{1\over6}\,\kappa\Theta s_au^a+ {1\over2}\,\kappa s_aA^a- \kappa s_{a(bc)}A^au^bu^c- \kappa s_{a\langle bc\rangle}u^a\sigma^{bc}- \kappa s_{a[bc]}u^a\omega^{bc}\,.  \label{ECRay1}
\end{eqnarray}
The above monitors the volume expansion/contraction of matter with nonzero spin within the framework of the Einstein-Cartan theory, with no restrictions on the nature of the matter fields involved. In fact, simplified versions of expression (\ref{ECRay1}) have been used to investigate the prevention of singularities in isotropic and anisotropic spacetimes with torsion (e.g.~see~\cite{K1}-\cite{SH}).

In the case of spatial anisotropy, one should also involve the shear propagation formula. By means of (\ref{algECFE2}a) and the Cartan field equations (see Eq.~(\ref{shev2}) in \S~\ref{ssSVE}), the latter reads
\begin{eqnarray}
h_{\langle a}{}^ch_{b\rangle}{}^d\dot{\sigma}_{cd}&=& - {2\over3}\,\Theta\sigma_{ab}- \sigma_{c\langle a}\sigma_{b\rangle}{}^c- \omega_{c\langle a}\omega_{b\rangle}{}^c+ {\rm D}_{\langle b}A_{a\rangle}+ A_{\langle a}A_{b\rangle}+{1\over2}\,\kappa h_{\langle a}{}^ch_{b\rangle}{}^dT_{cd}\nonumber\\ &&- C_{\langle a}{}^c{}_{b\rangle}{}^d u_cu_d- {1\over3}\,\kappa\Theta h_{\langle a}{}^ch_{b\rangle}{}^du^ms_{mcd}- \kappa A_{\langle a}h_{b\rangle}{}^cs_{c(dm)}u^du^m+ {1\over2}\,\kappa A_{\langle a}h_{b\rangle}{}^cs_c\nonumber\\ &&-{1\over2}\,\kappa s_cu^c\sigma_{ab}- \kappa\left(\sigma_{\langle a}{}^c+\omega_{\langle a}{}^c\right) u^dh_{b\rangle}{}^ms_{mcd}\,.  \label{ECshev}
\end{eqnarray}
This expression, which shows that the matter spin acts as a source of shear anisotropy, may also be used to probe its implications for the nature of a potential singularity. For example, one could pose the question of whether non-vanishing spin favours pancake-like or cigar-like singularities.

Matter with non-vanishing spin can also trigger vorticity and affect the rotational behaviour of the host spacetime. Indeed, expression (\ref{algECFE2}b) and the Cartan field equations transform the vorticity evolution formula (see Eq.~(\ref{vortev2}) in \S~\ref{ssSVE}) into
\begin{eqnarray}
h_{[a}{}^ch_{b]}{}^d\dot{\omega}_{cd}&=& -{2\over3}\,\Theta\omega_{ab}+ 2\sigma_{c[a}\omega_{b]}{}^c+ {\rm D}_{[b}A_{a]}+ {1\over2}\,\kappa h_{[a}{}^ch_{b]}{}^dT_{cd}- C_{[a}{}^c{}_{b]}{}^du_cu_d\nonumber\\ &&+{1\over3}\,\kappa\Theta h_{[a}{}^ch_{b]}{}^du^ms_{mcd}- \kappa A_{[a}h_{b]}{}^ms_{m(cd)}u^cu^d+ {1\over2}\,\kappa A_{[a}h_{b]}{}^cs_c \nonumber\\ &&-{1\over2}\,\kappa s_cu^c\omega_{ab}- \kappa\left(\sigma_{[a}{}^c+\omega_{[a}{}^c\right) u^dh_{b]}{}^ms_{mcd}\,.  \label{ECvortev}
\end{eqnarray}
Note the explicit spin terms on the right-hand side of the above, revealing how the latter can act as a source of spacetime rotation.

Expressions (\ref{ECRay1})-(\ref{ECvortev}) reveal the involved way the spin of the matter affects the kinematics of the host spacetime. This complication makes it difficult to extract quantitative results from the aforementioned relations, without first specifying the nature of the spin tensor. Nevertheless, qualitative conclusions are possible. For example, the role of the spin vector in all three of the above equations depends on the inner product $s_au^a$, which itself is decided by the relative orientation of the two vectors. Following (\ref{ECRay1})-(\ref{ECvortev}), the spin-vector effect acts in tune with that of the expansion when $s_au^a>0$ and against it in the opposite case. When the $u_a$-field is contracting, on the other hand, the situation is reversed. Finally, for purely spacelike spin vectors the impact of the above term is null.

The full kinematic description of an Einstein-Cartan universe, in the presence of torsion and spin, also requires the associated constraints. These are obtained from Eqs.~(\ref{con1})-(\ref{con3}) in an analogous way and lead to the following expressions for the scalar constraint
\begin{eqnarray}
{\rm D}_a\omega^a&=& A_a\omega^a+ {1\over4}\,\varepsilon_{abc}u_dC^{d[abc]}+ {1\over12}\,\kappa\Theta\varepsilon_{abc}s^{[abc]}+ {1\over4}\,\kappa\varepsilon_{abc}\sigma_d{}^{[a}s^{bc]d}- {1\over2}\,\kappa s_{a(bc)}\omega^au^bu^c\nonumber\\ &&-{1\over4}\,\kappa\varepsilon_{abc}u_dA^{[a}s^{bc]d}\,,  \label{ECcon1}
\end{eqnarray}
the vector constraint
\begin{eqnarray}
{2\over3}\,{\rm D}_a\Theta&=& {\rm D}^b\sigma_{ab}- {\rm curl}\omega_a- 2\varepsilon_{abc}A^b\omega^c- \kappa h_a{}^bT_{cb}u^c- {1\over3}\,\kappa\Theta h_a{}^{[c}h_b{}^{d]}s_{cd}{}^b- {1\over3}\,\kappa\Theta h_a{}^bs_b \nonumber\\ &&-\kappa h_a{}^{[c}\sigma_b{}^{d]}s_{cd}{}^b+ {1\over2}\,\kappa\sigma_{ab}s^b+ \kappa h_a{}^{[c}\varepsilon_b{}^{d]}{}_ms_{cd}{}^{[b}\omega^{m]}- {1\over2}\,\kappa\varepsilon_{abc}s^{[b}\omega^{c]} \nonumber\\ &&+\kappa h_a{}^{[c}A^{d]}s_{cdb}u^b  \label{ECcon2}
\end{eqnarray}
and the tensor constraint
\begin{eqnarray}
{\rm curl}\sigma_{ab}&=& -{\rm D}_{\langle b}\omega_{a\rangle}- 2A_{\langle a}\omega_{b\rangle}+ {1\over2}\,\varepsilon_{cd\langle a}C_{b\rangle m}{}^{[cd]}u^m- {1\over6}\,\kappa\Theta\varepsilon_{cd\langle a}h_{b\rangle m} s^{cdm}- {1\over2}\,\kappa\varepsilon_{cd\langle a}\sigma_{b\rangle m}s^{cdm}\nonumber\\ &&+{1\over2}\,\kappa\varepsilon_{cd\langle a}\sigma_{b\rangle}{}^cs^d- {1\over2}\,\kappa\varepsilon_{cd\langle a}\varepsilon_{b\rangle mn}s^{cd[m}\omega^{n]}- {1\over2}\,\kappa\omega_{\langle a}s_{b\rangle}+ {1\over2}\,\kappa\varepsilon_{cd\langle a} A_{b\rangle}u_ms^{cdm}\,.  \label{ECcon3}
\end{eqnarray}
It goes without saying that, in the absence of torsion and spin, Eqs.~(\ref{ECRay1})-(\ref{ECcon3}) reduce to their standard general-relativistic counterparts~\cite{TCM}.

\subsection{Raychaudhuri's equation in Einstein-Cartan
universes}\label{ssREE-CUs}
The mean kinematics of an Einstein-Cartan universe, with spacetime torsion and matter spin that obey the associated field equations (see \S~\ref{ssE-CFEs} earlier), are monitored by Raychaudhuri's formula (see Eq.~(\ref{ECRay1}) above). As stated in \S~\ref{ssRE}, positive terms on the right-hand side of (\ref{ECRay1}) tend to accelerate/declerated the expansion/contraction of the medium, while negative ones act in the opposite way. We should also note that, in the presence of torsion and spin, the scalar $T_{(ab)}u^au^b$ seen on the right-hand side of Eq.~(\ref{ECRay1}) is not necessarily positive, namely the weak energy condition does not always apply in Einstein-Cartan universes, even when dealing with otherwise conventional matter.

Raychaudhuri's formula can simplify considerably under certain conditions. For instance, when the fluid flow-lines are autoparallel curves, the 4-acceleration vanishes identically. If, in addition, the particle worldlines are irrotational and shear-free, expression (\ref{ECRay1}) reduces to
\begin{equation}
\dot{\Theta}= -{1\over3}\,\Theta^2- \kappa T_{(ab)}u^au^b- {1\over2}\,\kappa T- {1\over6}\,\kappa\Theta s_au^a\,.  \label{ECRay2}
\end{equation}
The last term on the right-hand side of the above vanishes when the spin vector is spacelike. When $s_a$ has a timelike component, on the other hand, the effect depends on whether the fluid is contracting or expanding (i.e.~on the sign of $\Theta$) and on the``tilt angle'' between the spin vector and the 4-velocity of the matter fields.

Alternatively, in the special case of totally antisymmetric torsion, we have $s_{abc}=s_{[abc]}$ and $s_a=0$ (see expression (\ref{CFE2}) in \S~\ref{ssE-CFEs}), in which case, the associated Raychaudhuri equation reads
\begin{equation}
\dot{\Theta}= -{1\over3}\,\Theta^2- \kappa T_{(ab)}u^au^b- {1\over2}\,\kappa T- 2\left(\sigma^2-\omega^2\right)+ {\rm D}_aA^a+ A_aA^a- \kappa s_{abc}u^{[a}\omega^{bc]}\,.  \label{ECRay3}
\end{equation}
Specifying the nature of the matter further, namely introducing an expression for the canonical spin tensor ($s_{abc}$) should generally allow one to evaluate the spin terms on the right-hand side of (\ref{ECRay2}), (\ref{ECRay3}) and thus estimate their effect on the mean kinematics of the fluid in question.

Perhaps the simplest case is the so-called Weyssenhoff fluid~\cite{WB}. This is a macroscopically continuous medium, which is microscopically characterised by the spin of the matter. The latter is monitored by the antisymmetric spin-density tensor ($s_{ab}=s_{[ab]}$), which is related to the canonical spin tensor by means of~\cite{OK}
\begin{equation}
s_{abc}= s_{ab}u_c\,,  \label{Weyss}
\end{equation}
while it satisfies the so-called ``Frenkel condition'', namely
\begin{equation}
s_{ab}u^b= 0\,.  \label{Frenkel}
\end{equation}
In other words, the spin-density tensor is spacelike in the rest-frame of the matter.\footnote{The presence of 3-dimensional antisymmetric second-rank tensor, which is essentially spacelike vector, defines a preferred spatial direction. This makes the Weyssenhoff fluid incompatible with the Cosmological Principle~\cite{Ts}.} The above conditions combine to ensure that the canonical spin tensor of Weyssenhoff-type media is trace-free by construction, which in turn guarantees that the canonical spin vector vanishes identically (i.e.~$s_a= s_{ba}{}^b=-s_{ab}{}^b=0$).\footnote{When dealing with Weyssenhoff-type media, the torsion and the spin-density tensors are related by $S_{abc}=-\kappa u_as_{bc}/2$, while the associated torsion vector vanishes (combine~Eqs.~(\ref{CFE1}), (\ref{CFE2}) and (\ref{Weyss}), (\ref{Frenkel}))} It should be noted, however, that, although the Weyssenhoff fluid provides a useful paradigm for studying the classical spin effects, it is of limited use from the field theoretical perspective (see~\cite{LP} for a discussion).

The canonical energy-momentum tensor of the Weyssenhoff fluid is that of an an ideal medium, of energy density $\rho$ and isotropic pressure $p$, with an additional contribution from the presence of spin. In particular, following~\cite{AKP}, we have
\begin{equation}
T_{ab}= \rho u_au_b+ ph_{ab}- A^cs_{ca}u_b\,,  \label{Wcemt}
\end{equation}
implying that $T_{(ab)}u^au^b=\rho$ and $T=3p-\rho$ (as a result of the Frenkel condition -- see Eq.~(\ref{Frenkel}) above). Therefore, on using the auxiliary relation (\ref{algECFE1}a), we have $R_{(ab)}u^au^b=\rho+3p$. This means that the spin of the Weyssenhoff fluid does not directly contribute to the local gravitational field

Applying (\ref{Weyss})-(\ref{Wcemt}) and the associated corollaries to Eq.~(\ref{ECRay3}) leads to the Raychaudhuri formula of an Einstein-Cartan spacetime filled with a Weyssenhoff-type medium, namely to
\begin{equation}
\dot{\Theta}= -{1\over3}\,\Theta^2- {1\over2}\,\kappa(\rho+3p)- 2\left(\sigma^2-\omega^2\right)+ {\rm D}_aA^a+ A_aA^a\,.  \label{WRay1}
\end{equation}
Formalistically, the latter is identical to its classical general-relativistic counterpart (e.g.~see \S~1.3.1 in~\cite{TCM}). Nevertheless, there are differences due to the presence of torsion and spin, which are revealed by appealing to the kinematic relations between a Riemann-Cartan and a purely Riemannian spacetime (see expressions (\ref{kin1}) and (\ref{kin2}) in \S~\ref{ssCvsRVs} earlier). When dealing with a Weyssenhoff fluid, the aforementioned relations reduce to
\begin{equation}
\Theta= \tilde{\Theta}\,, \hspace{19mm} \sigma_{ab}= \tilde{\sigma}_{ab}\,,  \label{Wkin1}
\end{equation}
\begin{equation}
\omega_{ab}= \tilde{\omega}_{ab}- {1\over2}\,\kappa s_{ab} \hspace{10mm} {\rm and} \hspace{10mm} A_a= \tilde{A}_a\,,  \label{Wkin2}
\end{equation}
respectively. Consequently, the kinematic variables of Weyssenhoff-type media are identical to their general relativistic analogues, with the exceptopn of the vorticity (see also~\cite{MT}). Moreover, starting from the definition of covariant differentiation, one can easily verify that $\dot{\Theta}=\tilde{\Theta}^{\prime}$ and ${\rm D}_aA^a= \tilde{\rm D}_a\tilde{A}^a$ (recall that primes and tildas indicate purely Riemannian environments). All these mean that the introduction of Weyssenhoff-type media modifies the standard Raychaudhuri equation solely through spin-induced effects to the rotational behaviour of the host spacetime.

The spin effects emerge after substituting the above given relations into the right-hand side of (\ref{WRay1}), which leads to the following version of the Raychaudhuri equation of a Weyssenhoff fluid
\begin{equation}
\tilde{\Theta}^{\prime}= -{1\over3}\,\tilde{\Theta}^2- {1\over2}\,\kappa(\rho+3p)- 2\left(\tilde{\sigma}^2 -\tilde{\omega}^2\right)+ \tilde{\rm D}_a\tilde{A}^a+ \tilde{A}_a\tilde{A}^a+ {1\over2}\,\kappa^2s^2- \kappa s_{ab}\tilde{\omega}^{ab}\,.  \label{WRay2}
\end{equation}
with $s^2=s_{ab}s^{ab}/2$ defining the magnitude of the spin-density tensor. The above expression reproduces the relation obtained in the (also 1+3~covariant) study of Weyssenhoff-type media given in~\cite{BHL}, when the differences in the metric signature and in the definitions of the vorticity and the spin-density tensors are accounted for. Our result also agrees with the familiar interpretation of an Einstein-Cartan spacetime filled with a Weyssenhoff fluid, as a Riemannian space containing a specific perfect fluid with nonzero spin.\footnote{Following~\cite{OK}, a perfect fluid with nonzero spin that satisfies conditions (\ref{Weyss}) and (\ref{Frenkel}) in a Riemann-Cartan spacetime, is equivalent to a general-relativistic medium with an effective stress-energy tensor of the form
\begin{equation}
T_{ab}= \left(\rho-{1\over4}\,\kappa s^2\right)u_au_b+ \left(p-{1\over4}\,\kappa s^2\right)h_{ab}- {1\over2}\,h^{cd}\tilde{\nabla}_d (s_{ca}u_b+s_{cb}u_a)\,. \label{Wefemt}
\end{equation}
Substituting the above into the classical Raychaudhuri equation leads to expression (\ref{WRay2}).} Note that in is common practise to assume that the microscopic spin orientation of the particles is random, in which case the macroscopic spin averages out to zero and one should only account for the quadratic spin contribution (i.e.~$\langle s_{ab}\rangle=0$, but $\langle s^2\rangle\neq0$). Then, the last term of (\ref{WRay2}) vanishes and the resulting expression agrees with the one obtained in~\cite{P}.

The quadratic spin-density term on the right-hand side of (\ref{WRay2}) inhibits the collapse or tends to accelerate the expansion of the Weyssenhoff fluid. Thus, spin and vorticity act in tune, which is intuitively plausible. There is an additional effect as well, through the coupling of these two sources, which can go either way. The effect of the spin is highlighted further if we momentarily adopt the familiar general-relativistic scenario of purely gravitational ``forces'' acting on an irrotational and shear-free perfect fluid with spin. Then, Eq.~(\ref{WRay2}) reduces to
\begin{equation}
\tilde{\Theta}^{\prime}= -{1\over3}\,\tilde{\Theta}^2- {1\over2}\,\kappa(\rho+3p)+ {1\over2}\,\kappa^2s^2+ \Lambda\,,  \label{WRay3}
\end{equation}
where we have momentarily reinstated the cosmological constant ($\Lambda$). Thus, qualitatively speaking, the spin term on the right-hand side of the above plays the role of an effective (positive) cosmological constant (when $s=\;$constant), or that of a quintessence field (when $s=s(t)$). We should note, however, that the spin contribution alone is rather unlikely to affect the late-time evolution of an ever expanding universe. Therefore, spin does not seem a likely substitute for dark energy, or capable of leading to an asymptotically de Sitter final phase (e.g.~as that described in~\cite{BG}). On the other hand, since the spin effects become stronger with increasing density, they could have dominated the early stages of the expansion, or the final stages of a recollapsing universe. More specifically, the inclusion of the spin could in principle allow for a geometrical description of inflation, without the need of scalar fields~\cite{P1}. Also, when dealing with the purely-gravitational collapse of matter with nonzero spin, expressions (\ref{WRay2}), (\ref{WRay3}) -- the latter with $\Lambda=0$ -- implies that the particle worldlines will not focus if $\kappa s^2>\rho+3p$, in which case the associated singularity (future or past) can be averted.

\section{Discussion}\label{sD}
To this day, the Einstein-Cartan gravity remains a viable theory and it is still experimentally indistinguishable from general relativity. In fact, Sciama expressed little doubt that, had the electron spin been discovered before 1915, Einstein would have included torsion in his theory. By abandoning the symmetry of the affine connection, Cartan demonstrated that its antisymmetric part, known as torsion, becomes an independent variable of the spacetime, together with the metric tensor. Macroscopically, the source of torsion is the intrinsic angular momentum (the spin) of the matter, in analogy with its energy density which gives rise to spacetime curvature.

Once it reemerged, primarily through the work of Kibble and Sciama in the late 1950s, the Einstein-Cartan theory has always maintained a level of attention, since it is probably the simplest and most straightforward classical extension of general relativity. Many studies, especially the earlier ones, looked into the implications of torsion and spin for singularity formation and in particular their avoidance. With very few exceptions, the available studies are centered around the Weyssenhoff fluid, namely an ideal medium with nonzero spin that satisfies the so-called Frenkel condition. The latter, however, makes the Weyssenhoff-type media incompatible with the Copernican Principle and therefore puts them at odds with the Einstein-Cartan analogues of the Friedmann universes. In addition, although it offers a valuable classical paradigm, the Weyssenhoff fluid is of limited use from the field theoretical perspective. Most of the studies also start by splitting their equations into a purely general-relativistic component supplemented by a torsion/spin part. Here, instead, we have not imposed any a priori restrictions on the nature of the matter fields, or on the relation between torsion and spin, until the very end of the analysis. As a result, our original kinematic formulae apply to a general imperfect fluid with nonzero spin, residing in spacetimes with arbitrary torsion.

The evolution and constraint equations have been derived in successive stages, first by incorporating the effect of spacetime torsion and then by including the spin itself. This was achieved by connecting torsion and spin through the standard Einstein-Cartan and the Cartan field equations. Nevertheless, given the generality of our study, alternative (i.e.~non-standard) relations between the aforementioned two entities may also be used. In the familiar case of a Weyssenhoff-type medium, we recovered the results of earlier studies, this time via an alternative (longer though more general) route. Moreover, the effect of the spin vector was found to depend (both qualitatively and quantitatively) on its orientation relative to the 4-velocity of the matter. The shear and the vorticity evolution equations also show how spin acts as a source of kinematic anisotropy and can be used to investigate its role, as well as that of torsion, in anisotropic spacetimes. Assuming, for example, that the particle spin is aligned along a given axis of symmetry (e.g.~along a shear eigenvector), one could look into its potential implications for the evolution of the host spacetime. The macroscopic effect of the intrinsic angular momentum of the matter on the fluid vorticity, as well as their combined action, can also be probed further. In addition, our work sets the basis for the study of perturbations in Einstein-Cartan cosmologies, but to proceed one needs to supplement the associated conservations laws, accompanied by expressions monitoring the Weyl field and and the 3-curvature in the presence of torsion and spin. This, however, goes beyond the scope of the present article. Here, our main aim was to provide the theoretical background for studying the kinematics of Riemann-Cartan and of Einstein-Cartan spacetimes in as general a way as possible. We did so by employing the 1+3~covariant formalism, thus extending the classic general relativistic studies to spacetimes with nonzero torsion and spin.\newline

\noindent \textbf{Acknowledgments:} We would like to thank Christian B\"{o}hmer, Nikos Chatzarakis, Friedrich Hehl, Damos Iosifidis and Tassos Petkou for helpful discussions and comments. KP acknowledges support from a research fellowship at CWRU, where part of this work took place. JDB is supported by the Science and Technology Facilities Council (STFC) of the United Kingdom.


\begin{thebibliography}{99}
\bibitem{C} E. Cartan, C.R. Acad. Sci. (Paris) \textbf{174}, 593 (1922); E. Cartan, Ann. Ec. Norm. Sup. \textbf{40}, 325 (1923); E. Cartan, Ann. Ec. Norm. Sup. \textbf{41}, 1 (1924); E. Cartan, Ann. Ec. Norm. Sup. \textbf{42}, 17 (1925).
\bibitem{Ki} T.W.B. Kibble, J. Math. Phys. \textbf{2}, 212 (1961); D.W. Sciama, in \textit{Recent Developments in General Relativity} (Pergamon+PWN, Oxford, UK) p.~415; D.W. Sciama, Rev. Mod. Phys. \textbf{36}, 463 (1964).
\bibitem{HK} F. Hehl and E. Kr\"oner, Z. Phys. \textbf{187}, 478 (1965); F. Hehl, Abb. Braunschweig. Wiss. Ges. \textbf{18}, 98 (1966); A. Trautman, Bull. Polon. Acad. Sci. \textbf{20}, 185 (1972); A. Trautman, Bull. Polon. Acad. Sci. \textbf{20}, 503 (1972).
\bibitem{BH} M. Blagojevic and F.W. Hehl, \textit{Gauge Theories and Gravitation} (Imperial College Press, London, 2013).
\bibitem{K1} W. Kopczynski, Phys. Lett. A \textbf{39}, 219 (1972); W. Kopczynski, Phys. Lett. A \textbf{43}, 63 (1973); A. Trautman, Nature \textbf{242}, 7 (1973); B. Kuchowicz, Gen. Rel. Grav. \textbf{9}, 511 (1978).
\bibitem{K2} W. Kopczynski, Phys. Lett. A \textbf{43}, 63 (1973); B. Kuchowicz, J. Phys. A: Math. Gen. \textbf{8}, L29 (1975); B. Kuchowicz, Astrophys. Space Sci. \textbf{39}, 157 (1976); J.M. Nester and J. Isenberg, Phys. Rev. D \textbf{15}, 2078 (1977); N.A. Batakis and D. Tsoubelis, Phys. Rev. D \textbf{26}, 2611 (1982).
\bibitem{SH} J. Stewart and P. Hajicek, Nature \textbf{244}, 96 (1973); F. Hehl, P. von der Heyde and G.D. Kerlick, Phys. Rev. D \textbf{10}, 1066 (1974); G.D. Kerlick. Ann. Phys. \textbf{99}, 127 (1976); G. Esposito, Nuovo Cim. B \textbf{105}, 75 (1990); N.J. Poplawski, Gen. Rel. Grav. \textbf{44}, 1007 (2012).
\bibitem{T} J. Tafel, Phys. Lett. A \textbf{45}, 341 (1973); A.J. Fennelly, J.P. Krisch, J.R. Ray and L.L. Smalley, J. Math. Phys. \textbf{32}, 485 (1991); K. Atazadeh, JCAP \textbf{06}, 020 (2014); M. Mohseni, Gen. Rel. Grav. \textbf{47}, 24 (2015).
\bibitem{MT} D.P. Mason and M. Tsamparlis, Gen. Rel. Grav. \textbf{13}, 123 (1981).
\bibitem{P} D. Palle, Nuovo Cim. B, \textbf{114}, 853 (1999).
\bibitem{KS} S. Kar and S. Sengupta, Pramana \textbf{69}, 49 (2007).
\bibitem{BHL} S.D. Brechet, M.P. Hobson and A.N. Lasenby, Class. Quantum Grav. \textbf{24}, 6329 (2007).
\bibitem{WB} J. Weyssenhoff and A. Raade, Acta Phys. Pol. \textbf{9}, 7 (1947).
\bibitem{E} G.F.R. Ellis, in \textit{General Relativity and Cosmology, Varenna Lectures}, Edit. R.K. Sachs (Academic Press, New York, 1971)~p.~104; G.F.R. Ellis, in \textit{Carg\`ese Lectures in Physics}, Edit. E. Schatzman (Gordon and Breach, New York, 1973)~p.~1.
\bibitem{P1} N.J. Poplawski, Phys. Lett. B \textbf{694}, 181 (2010).
\bibitem{S} L.L. Smalley, Phys. Rev. D \textbf{18}, 3896 (1978); M. Tsamparlis, Phys. Rev. D \textbf{24}, 1451 (1981).
\bibitem{HvdHK} F.W. Hehl, P. von der Heyde and G.D. Kerlick, Rev. Mod. Phys. \textbf{48}, 393 (1976); F.W. Hehl and Y.N. Obukhov, Annales Fond. Broglie \textbf{32}, 157 (2007).
\bibitem{Pa} A. Papapetrou, \textit{Lectures on General Relativity}, (D. Reidel Publishing Company, Dordrecht, Holland, 1974).
\bibitem{J} S. Jensen, \textit{General Relativity with Torsion} (University of Chicago, Chicago, IL, 2005).
\bibitem{P2} N.J. Poplawski, Astronom. Rev. \textbf{8}, 108 (2013).
\bibitem{TCM} C.G. Tsagas, A. Challinor and R. Maartens, Phys. Rep. \textbf{465}, 61 (2008).
\bibitem{HE} S.W. Hawking and G.F.R. Ellis, \textit{The large scale structure of space-time} (Cambridge University Press, Cambridge, UK, 1973); R.M. Wald, \textit{General Relativity} (University of Chicago Press, Chicago, 1984).
\bibitem{hL} H. het Lam, MSc Thesis, Utrecht University, The Netherlands (2016).
\bibitem{LP} S. Lucat and T. Prokopec, Class. Quantum Grav. \textbf{33}, 245002 (2016).
\bibitem{AKP} W. Arkuszewski, W. Kopczynski and V.N. Ponomariev, Ann. Inst. Henri Poincar\'e \textbf{21}, 89 (1974); C.G. B\"ohmer, Class. Quantum Grav. \textbf{21}, 1119 (2004).
\bibitem{OK} Y.N. Obukhov and V.A. Korotky, Class. Quantum Grav. \textbf{4}, 1633 (1987).
\bibitem{Ts} M. Tsamparlis, Phys. Lett. A \textbf{75}, 27 (1979).
\bibitem{BG} J.D. Barrow and G. G\"otz, Phys. Lett. B \textbf{231}, 228 (1989).
\end{thebibliography}
\end{document}